\documentclass[aps,showpacs,twocolumn,letterpaper]{revtex4}

\usepackage{graphics}
\usepackage{amssymb}
\usepackage[latin1]{inputenc}

\begin{document}

\title{ Logarithmic periodicities in the bifurcations of type-I 
intermittent chaos}
\author{Hugo L. D. de S. Cavalcante  and  J. R. Rios Leite}
\affiliation{Departamento de F\'{\i}sica,  Universidade Federal de Pernambuco,
50670-901 Recife, PE Brazil}
\date{\today}

\begin{abstract}
 The critical relations for statistical properties on saddle-node bifurcations
are  shown to display undulating  fine structure,
in addition to  their known smooth
dependence on the control parameter.
A piecewise linear map with the type-I intermittency is
studied and a log-periodic dependence is numerically obtained for the
average time between laminar events, the Lyapunov exponent and  attractor moments.
The origin of the oscillations is built in the
natural probabilistic measure of the map and can be traced back to the existence of
logarithmically distributed discrete values of the control parameter  giving Markov
partition.
Reinjection and noise effect dependences are discussed and indications are given on
how the oscillations are potentially applicable to complement predictions made with the usual
critical exponents, taken from data in critical phenomena.

\end{abstract}


\maketitle

Dynamical bifurcations, when the qualitative behavior of a natural phenomena
changes due to the variation of a control parameter, is a subject described by a
well founded mathematical theory \cite{Crawford}. Their extension from
equilibrium thermodinamic phase transitions to the bifurcations in non-equilibrium
systems is applied within and outside physics \cite{Manneville,Sornette}.
Some bifurcations are abrupt, like first order equilibrium transitions, and others
follow critical relations with a continuous
dependence in the control parameter. Sufficiently close to the bifurcation the
system properties follow characteristic functions giving rise to critical exponents
which are signatures of the phenomena.
Log-periodic oscillations modulating critical functional relations have been
reported in studies of bifurcations \cite{Sornette,Gluzman}.
As reviewed by Sornette \cite{Sornette},
 many ongoing research on
applications are active, among them the prediction of catastrophic events.
Transient chaos \cite{Grebogi-crise} and other escape phenomena \cite{Maier}
also display such log-periodicities.
In simple maps, the period doubling Feigenbaum cascade is an early example
where log-periodic behavior appears \cite{Derrida}.
In non attracting sets of two dimensional maps, the topological entropy have been
shown to present log-periodic oscillations \cite{Vollmer}.
%The subject of this letter is the log-periodic undulation on the dynamical bifurcation of
%the simplest one dimensional map showing intermittent chaos.

Among the dynamical bifurcation phenomena many can be cast on the class of
intermittent chaos, ranging from the onset of
turbulence \cite{Berge} to synchronism of chaotic
systems \cite{Zhan,Rosenblum-phase}.
 As originally proposed by Pomeau and Manneville
\cite{Pomeau,Berge,Manneville},
these instabilities can be modeled with
simple one-dimensional maps and classified as type-I, II, and III, according to the Floquet
multipliers of the map crossing the unity circle in the complex plane
at $1$, at a pair complex conjugate values or at $-1$, respectively.
Type I, which will be discussed here, occurs when a saddle-node or
tangent bifurcation is approached. Iterations of the map near the value of
the virtual fixed points are identified with the laminar events while the
reinjection iterates correspond to the turbulent bursts that occur in a
erratic manner.
Such bifurcation
has been reported in many experimental systems \cite{Jeffries,Lefranc}
and it is abundant in the logistic and other simple maps.
Renormalization group procedure to deal with type-I intermittency has already
been used \cite{Hirsch} and  give analytical asymptotic expressions for
statistical averages in the maps.
These are  treatments that only give smooth critical dependences.

The tangent bifurcation
%corresponds to a transition from order to chaos, and,
has analogy with second order thermodynamic phase transitions.
%and characteristic
%critical behavior observed in statistical properties as the control parameter
%is varied.
The continuous scale invariance of thermodynamical phase transitions
gives smooth monotonic critical relations for the statistical thermodynamical
variables  or the susceptibilities, as the order parameter approaches the
transition.
%What is not typical of thermodynamical phase transitions is the occurrence of
%fine structure oscillation in the value of the
However, as mentioned above, oscillations do occur in the corresponding
properties of non-equilibrium dynamical bifurcations.
Discrete scale invariance in the dynamical
systems bifurcations underly the oscillating features \cite{Sornette}.
In type-I intermittent chaos with the tangency having second and fourth order nonlinearity,
they appear with power-law periodicity \cite{Cavalcante01,Cavalcante03},
instead of log-periodicity.

Herein, to study the origin and properties of the log-periodic undulations,
the simplest unidimensional map exhibiting chaos and a saddle-node
bifurcation was considered. It is a piecewise linear map
first proposed to study critical
scaling laws of type-I intermittency by \citet{Shobu}.
A reason to take this map is the existence of an analytical approach
to find the discrete values of the control
parameter at which the scaled shape of the map function reproduces
itself.
The average length of laminar phases, the spectrum of fractal dimensions,
the Lyapunov exponent, and the
statistical moments of the chaotic variable, when calculated numerically,
show log-periodic oscillation.
Such numerical results correspond to time averages and
became possible due to the great computational power
achieved with current digital machines.
They can be compared to, and surpass in detail, analytical results from averages in ensembles
 valid at specific values of the control parameter.
The oscillation reported here are one such case.
% The period of the oscillation is shown to coincide with
%the  discrete scaling factor of the self replication.
%The coinciding values of time and ensemble averages indicate that the map is ergodic.
%Thus, the numerical results are comparable to analytical asymptotic relations
%and the period of the undulations is predictable.
Unfortunately we were unable to
extend the theory of renormalization group \cite{Hirsch} to predict such periodic property.

\begin{figure}[!hbtp]
\resizebox{6.5cm}{!}{\includegraphics{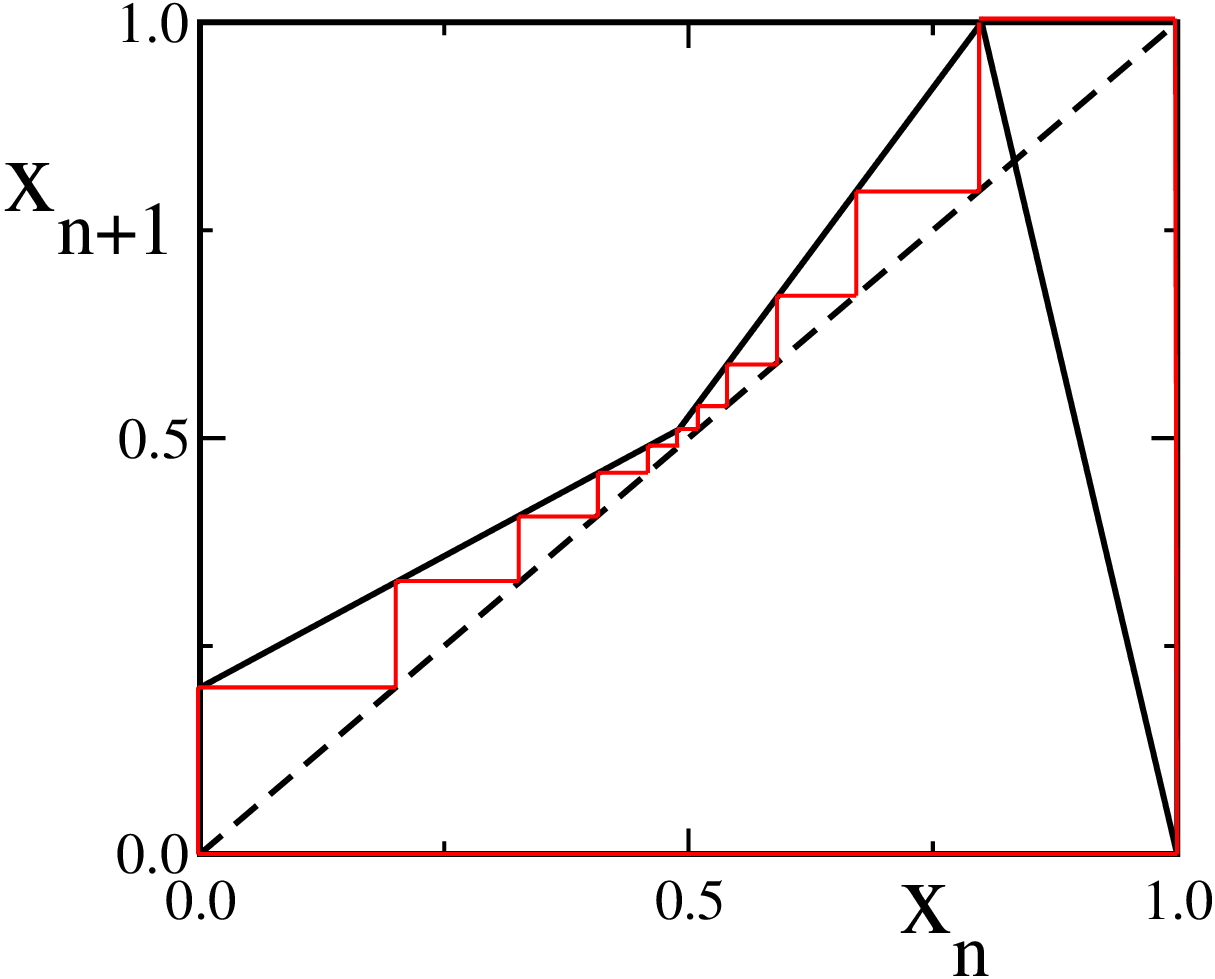}}
\caption{The So-Ose-Mori map with an unstable periodic orbit, calculated with
 $b=0.2$ and $\epsilon = 0.0332$. }
\label{fig:map}
\end{figure}

The equation for the map  is \cite{Shobu}
\begin{equation}
x_{n+1} =
        \left\{
                \begin{array}{ll}
                a x_{n} + b, & \mathrm{for}\ x_n \leq c \\
                a^{-1}[x_{n}-(1-b)]+1, & \mathrm{for}\ c < x_n \leq 1-b \\
                -b^{-1}(x_{n} -1), & \mathrm{for}\ x_n > 1-b
                \end{array}
        \right.
\label{eq:map}
\end{equation}
where $0<b<1/2$ is a constant
, $a= 1-2b +\epsilon>0$, and $c = (1-b)/(1+a)>0$
 are   expressed as functions
of the control parameter $\epsilon$.
The map is represented in figure~\ref{fig:map} with $b=0.2$ and
$\epsilon=0.0332$.
When $\epsilon <0$ the first two linear segments
cross the diagonal and a stable fixed point exists at the first crossing.
At $\epsilon =0$ the corner point of the first two linear segments
touch the diagonal, at $x=1/2$,
 and a bifurcation occurs similar to a saddle-node when two
fixed points disappear.
 For $\epsilon>0$, only the third linear segment intersects
the diagonal with an absolute value of slope greater than 1
and the map is chaotic.
Iterates at left of $(1-b)$ shall be considered as laminar, and
iterates on the negative slope will be referred to as reinjection
events.
%Any consideration of a narrower range around the corner point
%The first two linear segments change according $\epsilon$
%crossing the diagonal at $x=1/2$ for $\epsilon=0$.
%three linear continuous segments mimic the nonlinear
%normal form of quadratic
With  $\epsilon>0$,
$c$ is approximately proportional to
the minimum distance between the map and the $y=x$ diagonal.

\citet{Shobu} have shown that, for this map, simple Markov partitions
exist at discrete values of $\epsilon$.
%A family of orbits whose
%iterates are the limits of the intervals of the
%simplest Markov partition of the map are shown to appear
%at discrete
%values of $\epsilon$.
When this happens, as one iterates the map, the probability of visits,
or the natural measure, over the unit interval
is a countable set of step intervals with constant value.
The map dynamics is that of a
finite  Markov chain, diverging for $\epsilon \rightarrow 0$,
and it is analytically solvable.
The appearance of unstable periodic orbits, reinjecting at $x=0$ and iterating $2n+2$
times until reaching precisely the value $1$, mark the condition
for the Markov partition with $2n+1$ constant steps for the
natural measure. Figure~\ref{fig:map} shows the map and the iterates of
such an orbit with period 12.
The specific values of $\epsilon = \epsilon_{n}$ for which the
orbit of period $2n+2$ exists follow the recursive
relation $\epsilon_n = (1-2b)\epsilon_{n-1}$  \cite{Shobu},
and, as good approximation for $n>3$,
\begin{equation}
\epsilon_n = 2b\left(\frac{1-b}{1+b}\right)(1-2b)^{n-1}.
\label{eq:epsilonn}
\end{equation}
% is always positive for $\epsilon>0$.
%and thus, they accumulate most of the density of probability of visitation
%as the local Lyapunov for every other unstable orbit is greater than zero.
%%%as good approximation for $n>3$.
%Using the expression for $n<3$ leads to a
%wrong count on the number of orbits and wrong values of $\epsilon_n$ giving
%the Markov partitions.
The factor $(1-2b)$ will appear as the period of the fine structure in
logarithmic scale.
At the values of $\epsilon_n$ one can obtain analytically the ensemble averages
for the average length of laminar events, $\left<l\right>$,
 for the first order moment of the attractor, $\left<x\right>$, \cite{Shobu},
and for the Lyapunov exponent, $\left<\lambda\right>$,
%which will be taken without braces as $\lambda$.
The analytical expressions first determined by \citet{Shobu},
and resulting from the summation with the stepwise uniform probabilistic measure, are
\begin{equation}
\left< l \right> \cong  \frac{1}{\ln(1-2b)}\ln\left(\frac{1}{\epsilon_n}\right)\ ,
\label{eq:laven}
\end{equation}
\begin{equation}
\left<x\right> \cong  \frac{1}{2} + \frac{1}{2n}\ ,
\label{eq:aven}
\end{equation}
and
\begin{equation}
\left<\lambda \right> \cong  \frac{1}{n}
	\left[\ln\frac{1}{b}+\frac{1}{(b-2)}\ln\frac{1}{(1-2b)}\right].
\label{eq:lambdan}
\end{equation}

The ensemble averages must coincide with time averages $\overline{l}$,
$\overline\lambda$ and
$\overline{x}$, calculated using iterations of the map and starting
from typical initial conditions.
The length of a laminar event is considered as the number of
iterations ``in the channel'' before a reinjection.
%Such channel has been defined as the two
%leftmost linear segments in the map.
Any consideration of a narrower range around the corner point as the channel
causes just a shift in the value of $\overline{l}$ for a given $\epsilon$,
but the slope and fine structure dependence on $\epsilon$ remains the same.
The Lyapunov exponent (we use $\lambda$ for $\overline\lambda$ and$\left<\lambda\right>$)
 was calculated numerically using
a large number of iterations, $N$, with the expression
\begin{equation}
\lambda = %\lim_{N\rightarrow \infty}
\frac{1}{N} \sum_{n=1}^{N} \ln
\left| f^{\prime}(x_n) \right|,
\label{eq:lyapunov}
\end{equation}
and the average of the dynamical variable
$\overline{x}$  with
\begin{equation}
\overline{x}=%\lim_{N\rightarrow \infty}
\frac{1}{N} \sum_{n=1}^{N}x_n.
\label{eq:ave}
\end{equation}

\begin{figure}[bthp]
%\resizebox{8cm}{!}{\includegraphics{ave-l-linear-epsn.eps}}
\resizebox{8cm}{!}{\includegraphics{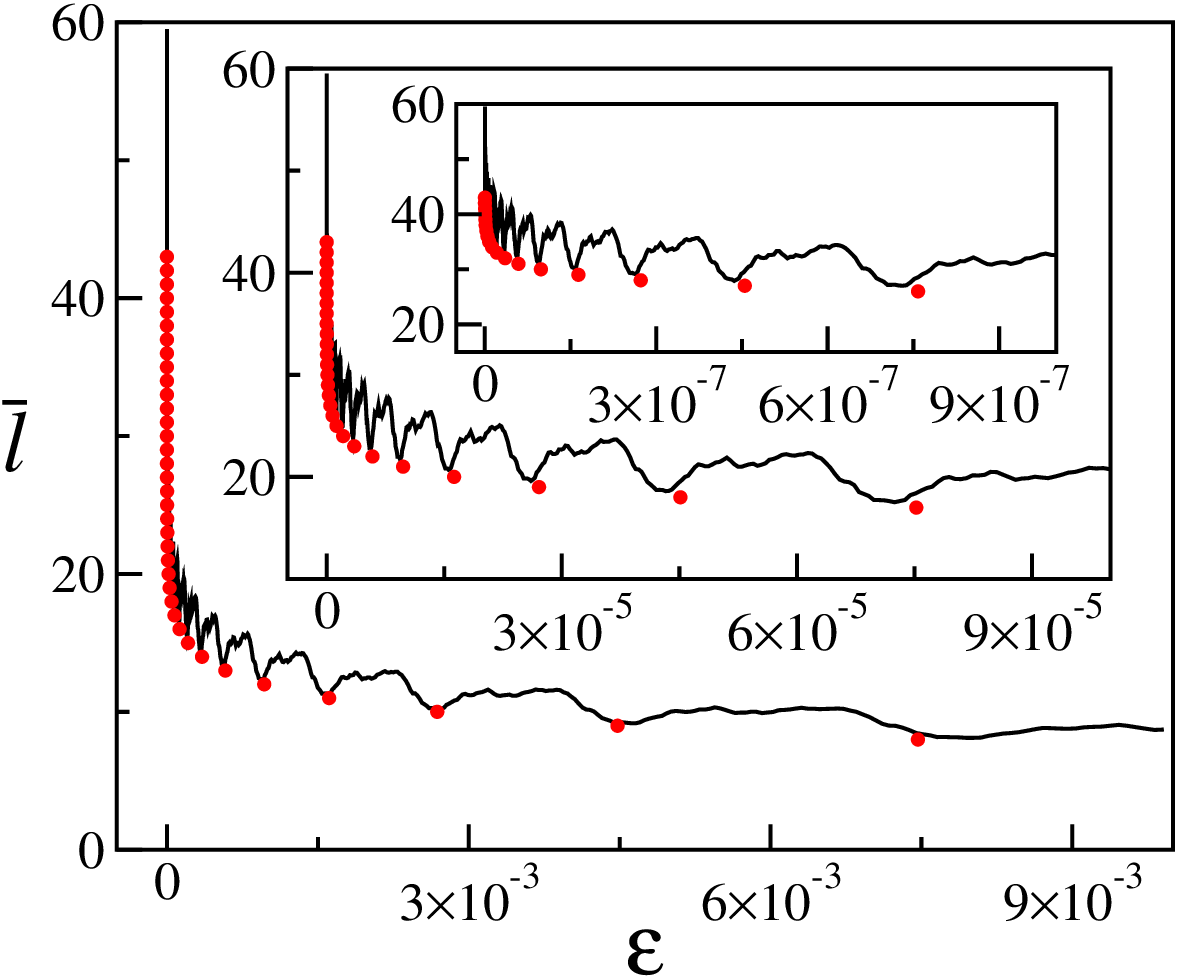}}
\caption{Average length of laminar events. The dots are analytical ensemble averages
%calculated analytically
%at discrete values of control parameter $\epsilon$
% using the probability measure
% from ref.~\cite{Shobu}
and the continuous curves are  numerical time averages,
iterating the map,
 as discussed in text. }
\label{fig:average-l}
\end{figure}

Figure~\ref{fig:average-l} shows the average length of laminar events
calculated both with time averages and ensemble average.
%with the probabilistic measure
  In the last case, only a discrete set
of values of $\epsilon$ from eq.~\ref{eq:epsilonn} can be done.
The values of $\overline{l}$ decrease with the control parameter $\epsilon$.
Thus if one calculates and  interpolates the analytical discrete
values \cite{Shobu} one never gets the periodic effects.
Conversely, for the time averages
the values of $\epsilon$ can be picked at will.
In the figure for $\overline{l}$,  2000 values of the control parameter $\epsilon$ were equally
spaced on a logarithmic scale and  at each one,  $N=3\times 10^{7}$ iterations
were done for the averages, with double precision floating point arithmetics.
One observes clearly that, in addition to the smooth power-law divergence as
$\ln\epsilon^{-1}$, $\overline{l}$ is decorated with
a modulation whose period is given by the positions at which each Markov
partition is formed.

\begin{figure}[htbp]
\resizebox{8cm}{!}{\includegraphics{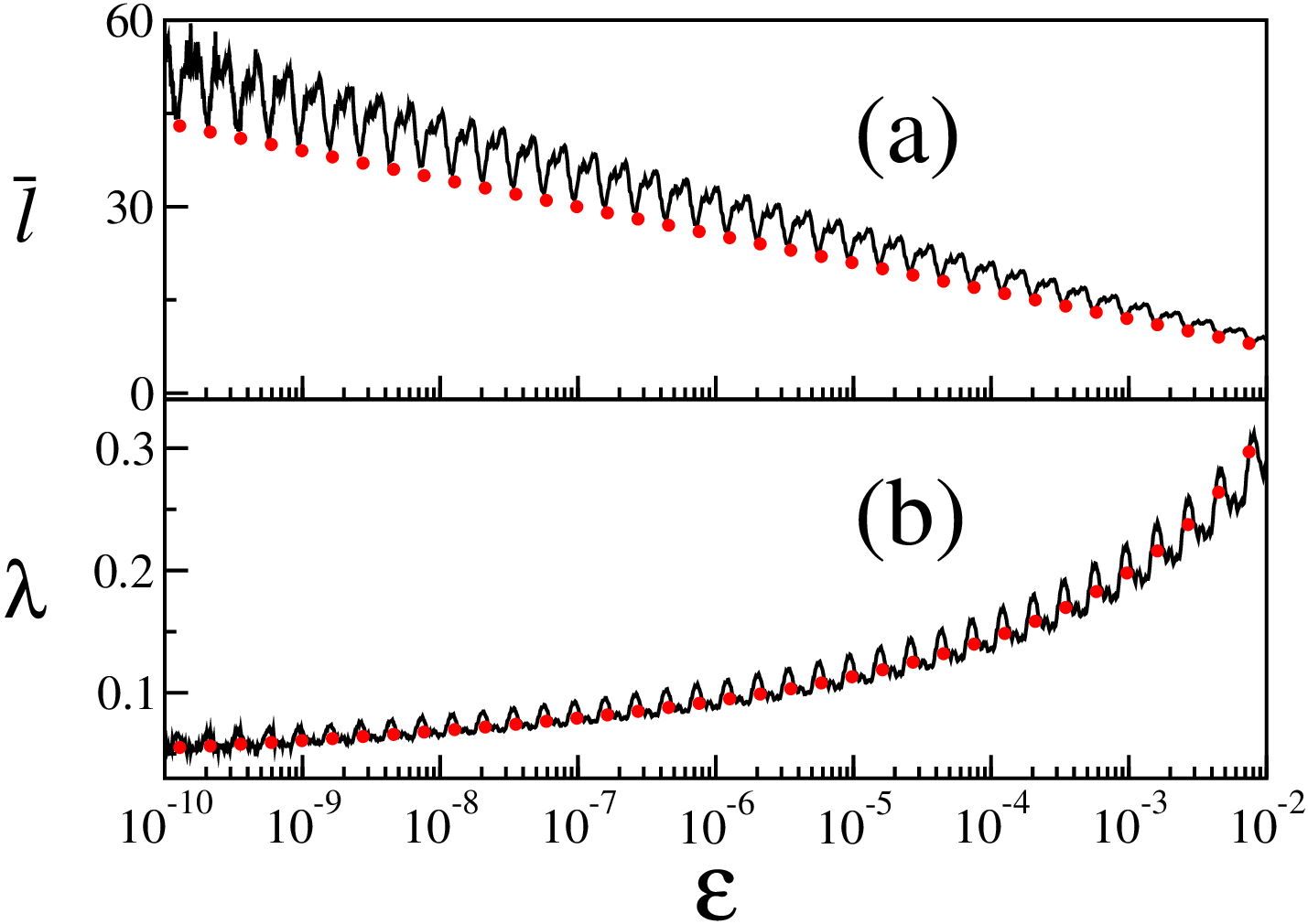}}
\caption{
%Averages in the noiseless map: (a)
Average length of laminar events (a),
and (b) average Lyapunov exponent, $\lambda$, in the SOM map.
 The dots are analytical ensemble averages
%calculated analytically at discrete values of control parameter $\epsilon$
%using the probability measure from ref.~\cite{Shobu}.
 and the continuous curves are numerical time averages.
 %, iterating
%the map as discussed in text.
}
\label{fig:logperiodic}
\end{figure}

\begin{figure}[htbp]
\resizebox{8cm}{!}{\includegraphics{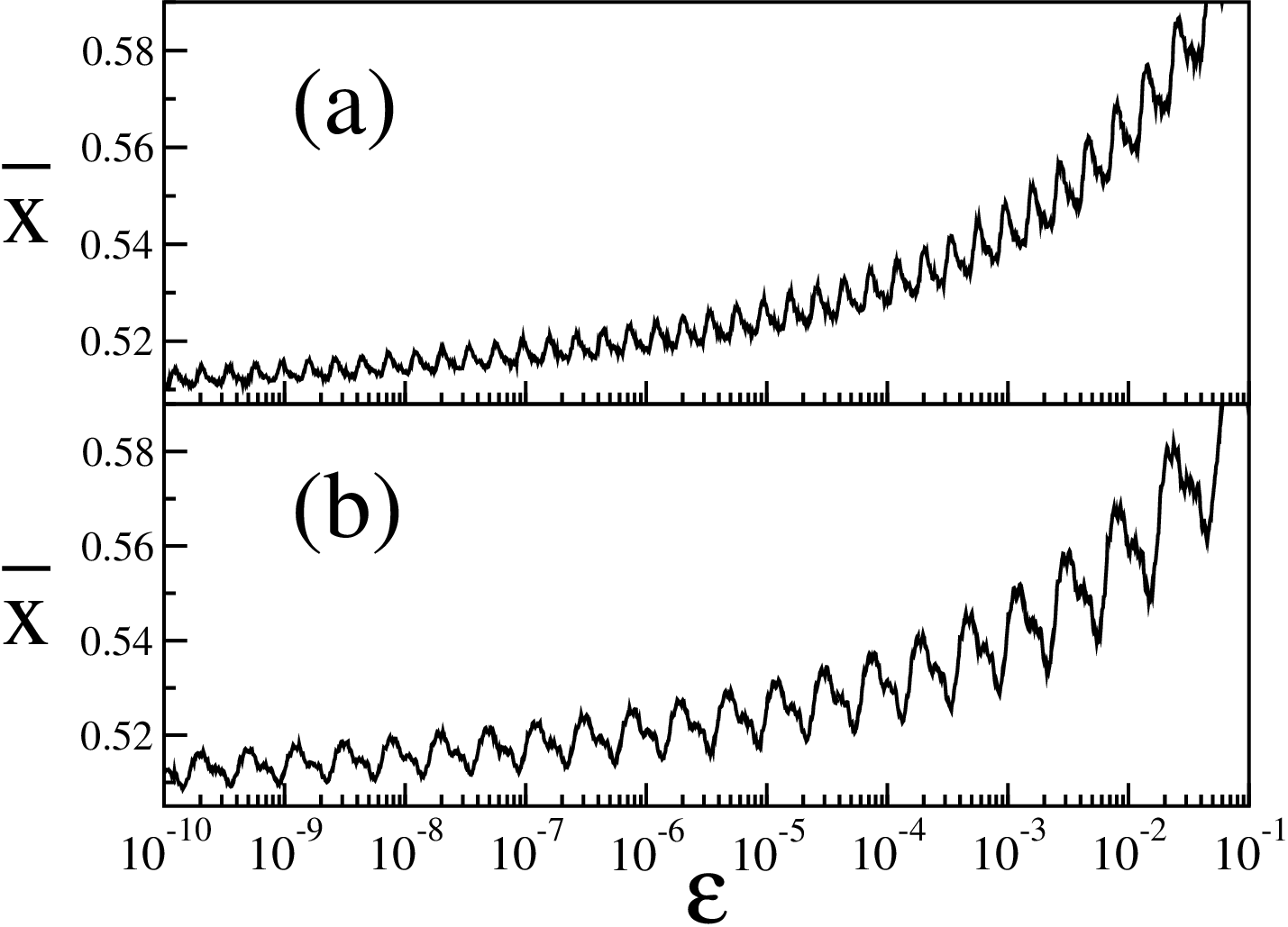}}
\caption{Average of the dynamical variable in the SOM map with
different scaling factor: (a) $b=0.2$ and (b) $b=0.3$.
%The periodicity of the fine structure changes as $\ln(1-2b)$
}
\label{fig:avex-b}
\end{figure}

Numerical computations of the average length of laminar events and average
Lyapunov exponent are show in fig.~\ref{fig:logperiodic}.
Logarithmic scale is used in the absciss and
the critical behavior of $\overline{l}$ has a linear smooth envelope behavior as
$\ln\epsilon^{-1}$ for small $\epsilon$.
%so exponent $-1$  for
%$\overline l$ shows as a linear dependence for the envelope.
  Furthermore, its
 oscillations are periodic
 as function of $\ln\epsilon$. The behavior for $\left<x\right>$ is proportional
to the one exhibited by $\lambda$, as one expects comparing
eqs.~\ref{eq:aven} and~\ref{eq:lambdan}.
Changing $b$ on equation \ref{eq:map} changes the
spacing between successive $\epsilon_n$, and the corresponding period obtained
in time averages follows consistently
 with  the scaling factor $\epsilon_n / \epsilon_{n+1} = (1-2b)$.
 Therefore, numerical artifacts for
log-periodicity are completely ruled out in the computations.
 This $b$ dependency is shown in fig.~\ref{fig:avex-b}.
 The same oscillating features were obtained (not show here)
for $D_q(\epsilon)$, the fractal dimension
of the attractor, calculated for $q=0,1$ and $2$.
 The critical exponent of the smooth
envelope of any of the averages do not change with $b$.

%Most important,
%The undulations contains novel information about the system.
%not accessible with the smooth
%critical dependency approximation.

Calculations were also done using a random reinjection value in a map with
the  two first linear segments as the laminar phase part of  eq.~\ref{eq:map}.
% Figure~\ref{fig:averandom} shows the average
%of the variable with this random reinjection.
As long as the reinjection is non singular, oscillations with the same period
are always obtained. This fact is indication of the local nature of the
oscillations.
While the Markov partitions of the SOM maps are a global result,
their relevant contributions  for statistical averages near criticality come from
the uniform density steps near the narrow channel.
Any approximately uniform reinjection
on this region suffices to give oscillating averages.
%\begin{figure}[htbp]
%\resizebox{8cm}{!}{\includegraphics{avex-random.eps}}
%\caption{Average of the dynamical variable in the SOM map with random
%reinjection. The periodicity of the fine structure is the same as the
%deterministic reinjection case.}
%\label{fig:averandom}
%\end{figure}

\begin{figure}[h!btp]
\resizebox{8cm}{!}{\includegraphics{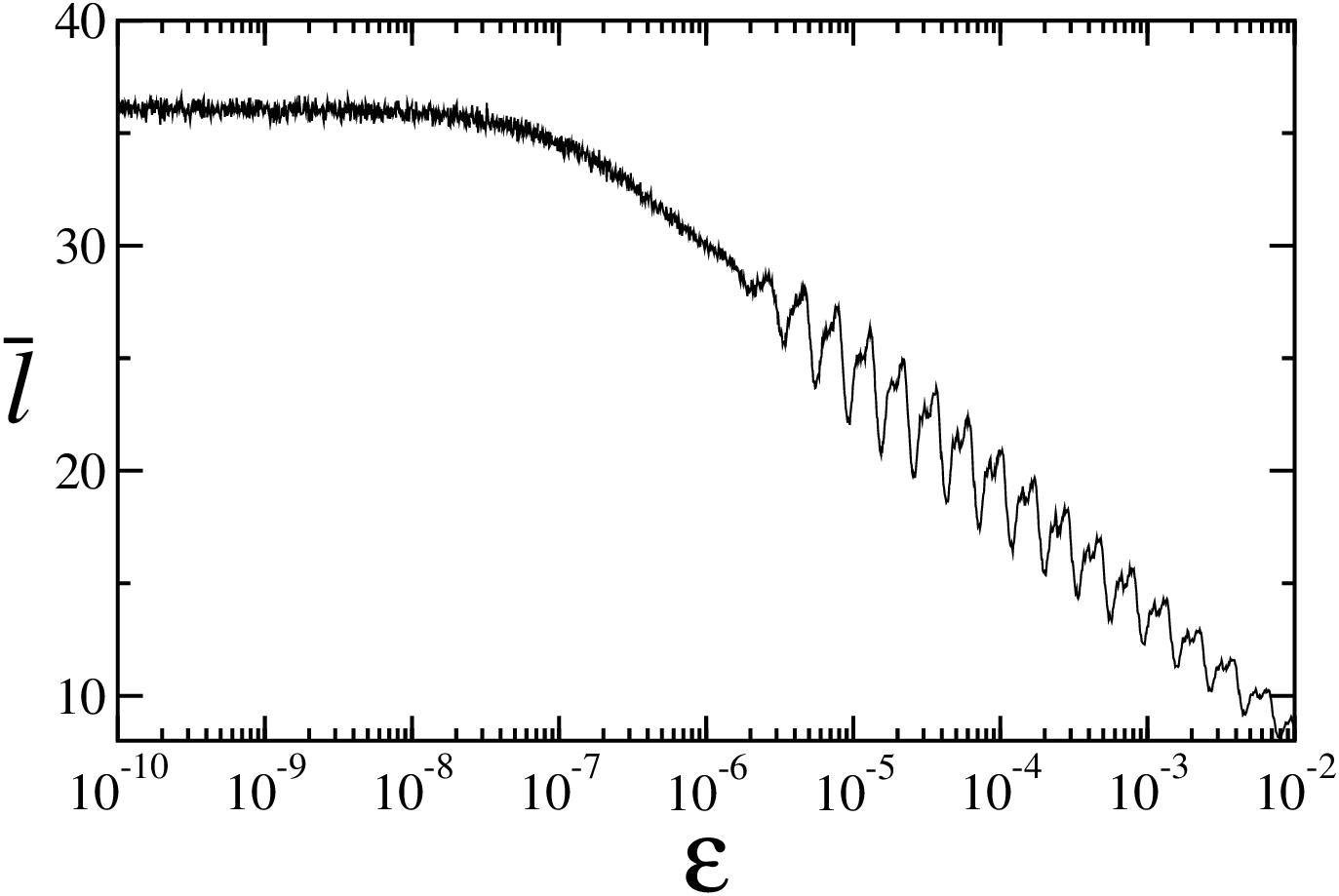}}
\caption{Averages length of laminar events in the SOM
%So-Ose-Mori
map with additive noise.}
\label{fig:avelnoisy}
\end{figure}

\begin{figure}[htbp]
\resizebox{8cm}{!}{\includegraphics{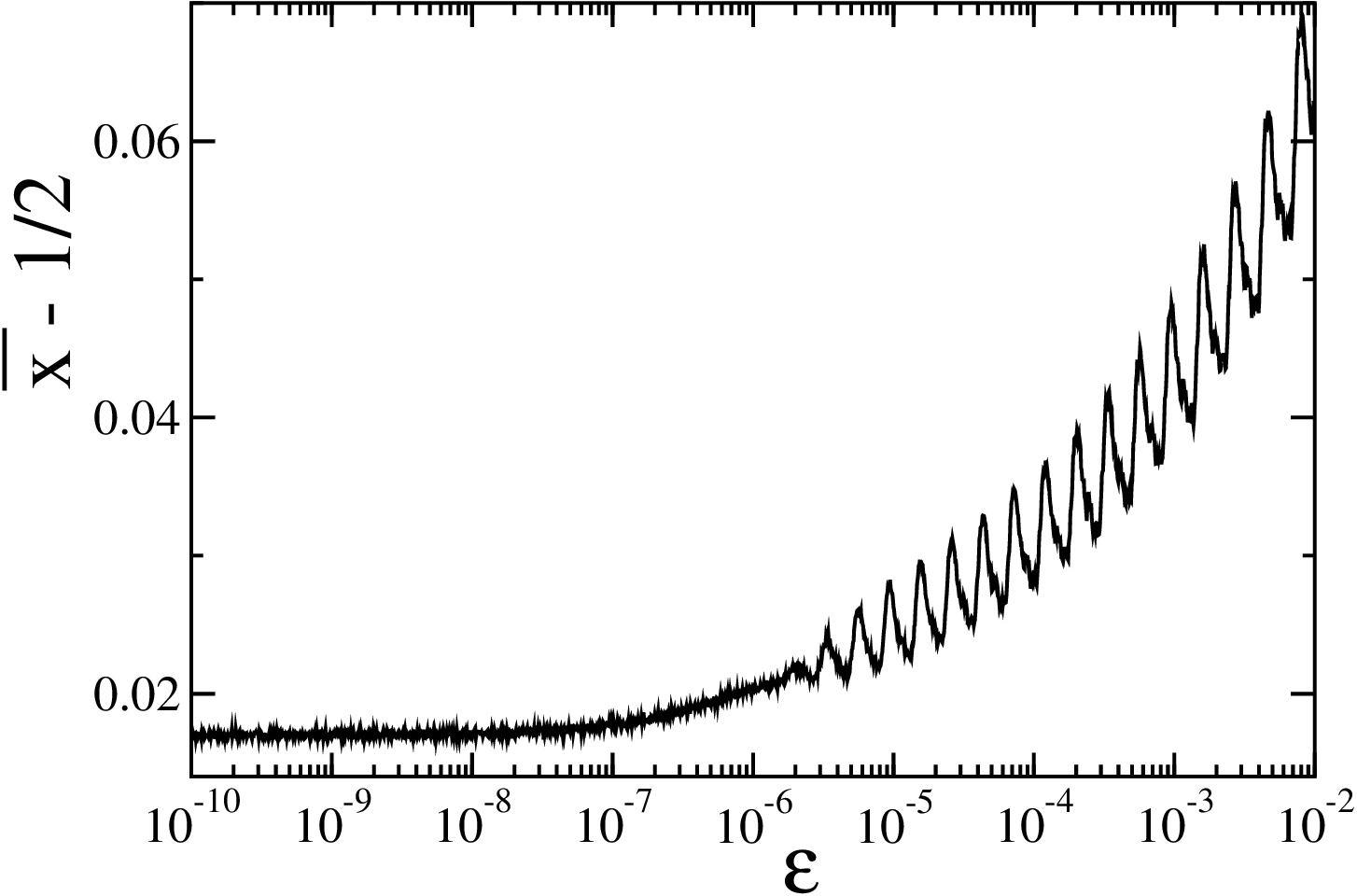}}
\caption{Average of the dynamical variable, $\overline x$, in the SOM map with
additive noise.}
\label{fig:avexnoise}
\end{figure}

Finally some discussion on noise.
The effects of noise in type-I intermittency have been studied
analytically \cite{Hirsch} and experimentally \cite{Cho}.
Here, a term of additive gaussian white noise with
amplitude $D$ was added to the right hand side of eq.~\ref{eq:map}.
 Fig.~\ref{fig:avelnoisy}  shows $\overline{l}$,
% the average length of laminar events
 when the noise amplitude was $D=10^{-6}$.
For $ \epsilon \gg D$ one expects the dynamics to remain unaffected.
When the separation between successive $\epsilon_n$ becomes of the order of
the noise amplitude, $D$, the corresponding and smaller oscillations are
blurred out.
 For even smaller $\epsilon$ ($\lesssim D$),
the number of iterations in the narrowest part of the channel
no longer depends on $\epsilon$,
because  a single noise event may cause a jump across the waist of the channel, which,
otherwise would have a number of iteration proportional to
$\ln\epsilon^{-1}$.
Therefore, $\overline{l}$ saturates.
%%%%%is added in the numerical iterations of eq.~\ref{eq:map}.
%The average of the dynamical variable
Such noise sensitivity is shown for $\overline{x}$ in fig.~\ref{fig:avexnoise}.
It is relevant to observe that, consistent with the above explanation, in
both figures, \ref{fig:avelnoisy} and \ref{fig:avexnoise},
the oscillations cease to exist for $\epsilon$ one order of magnitude higher
than the value
at which  the  slope of the average saturates.
One also may see a smoothing of the ultrafine structures on the undulations.
Smoothing,  erasure of the
undulations, and saturation of the amplitude envelope also occurs  in the
Lyapunov exponent  when noise is present.
 It may also be observed in the figures  that the
amplitude of the undulations on $\overline l$ and $\overline x$ are  bigger at  different
ranges of $\epsilon$. This might be used to justify the importance of acquisition
 of different data series and different  averaged quantities in experiments.

To conclude it must be stressed that oscillations in
the critical behavior of statistical properties occur universally in
 the tangent bifurcations of unidimensional normal form maps
 \cite{Cavalcante03} and in the logistic map \cite{Cavalcante01}. They also occur in
  the saddle-node bifurcations of continuous chaotic systems, as we verified numerically
  integrating the R\"ossler, the Lorenz and the
  equations for a laser with saturable absorber \cite{hugo-laser}.
   However none of these follow the log-periodic dependence.
   The map herein was chosen to stress
 the relationship between log-periodic fine structures, existence of Markov partition and
  self similar scalling. From a mathematical point of view
  much more remains to be done on this connection.
The undulations are important features of bifurcations that
can be subject to experimental tests and applications.
To the present, in  experiments with intermittent chaos only the smooth critical
behavior have been studied. The observation of log-periodic oscillations has
been proposed as a technique to anticipate the criticality in natural
phenomena, particularly in earthquakes \cite{Sornette}.
The simple map presented here is far from modeling earthquakes but it
is an workable example to show how the critical value of a
parameter controlling bifurcations might be predicted  ahead of
criticality, by  complementary fitting of
oscillations in the fine structure at higher values of the control
parameter, along with the usual fitting of the monotonic critical behavior.
It also shows how improvement in extracting information on a bifurcation with
intermittent chaotic should be enhanced by measuring different statistical variables on different
ranges of the control parameter.

\acknowledgments{Work partially supported by Brazilian
Agencies: Conselho
Nacional de Pesquisa e Desenvolvimento (CNPq) and
Financiadora de Estudos e
Projetos (FINEP)
}

\bibliography{som}

\end{document}